\documentclass[11pt, oneside]{article}   	
\usepackage{geometry}                		
\usepackage{graphicx}				
\usepackage{url}
\usepackage{latexsym}
\usepackage{amsmath}
\usepackage{amsthm, amsfonts}
\usepackage{algorithm} 
\usepackage{algpseudocode}  
\usepackage{multicol}
\usepackage{mathtools}
\usepackage{bm, float}
\usepackage{tikz}
\usepackage{tikz-qtree}
\usepackage{etoolbox}
\usepackage{hyperref}
\usepackage{subfigure}

\begin{document} 

\title{Layered Dirichlet Modeling to Assess the Changing Contributions of MLB Players as they Age}
\author{
Monnie McGee$^1$, Jacob Turner$^2$, and Bianca Luedeker $^3$\\
$^1$Southern Methodist University\\ $^2$Stephen F. Austin State University \\ $^3$Northern Arizona University}

\maketitle
\begin{center}
{\bf Abstract}
\end{center}
The productive career of a professional athlete is limited compared to the normal human lifespan. Most professional athletes have retired by age 40. The early retirement age is due to a combination of age-related performance and life considerations. While younger players typically are stronger and faster than their older teammates, older teammates add value to a team due to their experience and perspective. Indeed, the highest--paid major league baseball players are  those over the age of 35. These players contribute intangibly to a team through mentorship of younger players; however, their peak athletic performance has likely passed. Given this, it is of interest to learn how more mature players contribute to a team in measurable ways. We examine the distribution of plate appearance outcomes from three different age groups as compositional data, using Layered Dirichlet Modeling (LDM). We develop a hypothesis testing framework to compare the average proportions of outcomes for each component among 3 of more groups. LDM can not only determine evidence for differences among populations, but also pinpoint within which component the largest changes are likely to occur. This framework can determine where players can be of most use as they age.

\section{Introduction}

In professional baseball, the highest paid players are typically those over the age of 35 \cite{forbes}. Perhaps some additional pay is due to their popularity; however, it is unlikely that management would continue to pay large sums of money for a player unless it is perceived that the player is making a contribution to the overall winning percentage of the team. Some of the contribution is likely intangible, such as leadership and mentorship of younger players. Other contributions directly affect game outcomes. For example, there is evidence older players trade off their earlier ability to reach base via foot speed for an ability to reach base via outcomes that require more patience or fortitude, such as walks and hit by pitch \cite{bradbury}.  In this article, we propose a method to determine differences in the distribution of at-bat outcomes, which we model as a composition of six possible at-bat outcomes. We propose a likelihood ratio test for $G > 2$ groups, and apply the test for MLB baseball players near the beginning, middle, and end of their careers. Our analysis corroborates earlier analyses that show the shift in contributions of older players from more explosive outcomes to outcomes that require patience and experience.

It is well-known that physical maturation affects athletic performance \cite{mangine}, and it is further accepted that athletic performance peaks at a certain age, typically in the mid to late 20's, and then declines \cite{schulz}. Physical characteristics such as speed, strength, lung capacity, and flexibility, are harder to maintain as the body ages, and all of these are related to performance in professional baseball and other sports \cite{bradbury}. Studies have examined the trajectory of performance-related variables as players age, and all come to the same conclusion: by the time an athlete is 40 years old, most of the markers of athleticism have diminished. However, decline in athleticism can be mitigated by other factors. For example, a study of approximately 34,000 athletes in the Swedish Veteran Athletics data base, which covers track and field events over 120 years \cite{swedes}, showed that athletes who started when they were older and at a high performance level tended to experience the slowest rate of decline in performance \cite{geroscience}. For MLB players, early specialization in baseball has been linked to less longevity due to upper extremity injuries \cite{confino}; however, a recent review of baseball injuries showed that players can return to play with no discernible difference in performance \cite{swindell}. Some studies suggest that peak performance, as well as injury prevalence, interact with the type of activity. More explosive activities requiring strength and speed tend to favor younger athletes, while lower impact activities requiring excellent technique tend to favor more mature athletes \cite{gabbard}. 

The experience-productivity relationship for baseball, specifically, has been the subject of several articles. For example, \cite{krohn} measured a baseball player's lifetime batting average using a regression equation including variables for age and age-squared for each of the years that a batter remained in the major leagues. The model also included an error term which consisted of a variable for talent, which is time-invariant, and a time-varying individual performance term \cite{krohn}. The aim was to produce a career batting average for each player that was adjusted for age, talent, and experience. This model considered only batting average, and not individual at-bat outcomes. Hakes and Turner (2009) examine the relationship between salary of MLB players and their productivity as measured by on-base percentage plus slugging average (OPS) for 1985-2005. They divide MLB players into quintiles based on the indexed OPS level of a player's third-best season of more than 130 at bats. Instead of age as a predictor in their model, they use years of experience.


Bradbury (2008) defines athleticism for both hitters and pitchers. For hitters, athleticism is defined by a linear regression model with variables for batting average, walks per-plate appearance, double plus triples per at bat, home runs per at bat, on-base percentage, on-base plus slugging percentage, slugging average, and adjusted linear weights. Linear weights is a measure which is a weighted combination of singles, doubles, triples, home runs, walks, hit-by-pitch, stolen bases, caught stealing, and at-bat hits \cite{thorn}. The goal is to measure the age at which players peak in their performance according to these measures. Commensurate with most research on performance and aging, the peak age is typically between 27 and 29 for all measures of hitting performance \cite{bradbury}.

Most of the previous work has been to examine age at peak performance, and most of that work used linear regression and one selected measure of performance to determine that peak. We seek to examine the contribution of {\it all} offensive plate appearance outcomes of baseball within the context of a game. We treat each outcome as a component of a discrete distribution.  The statistical model we utilize to determine changes in age-related performance is based on the nested Dirichlet distribution (NDD) \cite{null2009}. We show that there is evidence that the proportions of plate appearance offensive outcomes change for different age groups of batters, and we further show exactly where the largest discrepancies in the outcomes lie. This ability to determine the component with the most impact sets our methodology apart from others.

This paper is structured as follows. In Section 2, we introduce the baseball outcome data and explore it with descriptive statistics and graphics. We show that plate appearance outcomes can be considered as individual components that make up a hitter's performance. To model these components as a whole, we introduce the nested Dirichlet distribution (NDD) in Section 3 and discuss estimation of its parameters. In section 4, we outline a hypothesis test to examine the similarity of the component structure across multiple groups, where each group has a nested Dirichlet distribution. We call this methodology {\it Layered Dirichlet Modeling}. We also include simulation results giving the Type I and Type II error properties of layered Dirichlet modeling (LDM) for various sample sizes and effect sizes.  In section 5, we apply LDM to examine the contribution of plate appearance outcomes for batters in three different age groups. We show that there is evidence for a difference in the mean proportion of outcomes for different age groups, which has implications for decisions about the batting order for a given team. The paper concludes in Section 6 with a discussion of potential extensions of this framework to factorial experimental designs, and to incorporate temporal dependence across athletic seasons. While the NDD framework is interesting and useful for analysis of composition data in its own right, our application here indicates the value in having diverse age groups of players on a team. We show that players of different ages contribute to the run production of a baseball team in important ways that have not previously been elucidated.


\section{The Data}

Baseball is a team sport that can be summarized as a sequence of ``battles'' or ``at bats'' between the pitcher and batter.  Each plate appearance results in a discrete outcome \cite{baumer23}. The outcome of any pitch is either a strike or a ball. Once the ball is hit, play continues until there is either an out or a batter reaches a base safely. Then, play stops until the next pitch.  Officially, an at-bat requires the batter to reach base via a hit, a fielder's choice, or an error. Here, we consider plate appearances, which is a completed turn as a batter, regardless of the result. In contrast, basketball and football are continuous-time games where discrete outcomes can occur (2 points, 3 points, or no basket in basketball; goal or no goal in football), but the various outcomes take place in continuous time.

In the case of baseball, 14 discrete outcomes are possible for each plate appearance. They are given in Table \ref{tab:components}, grouped by fly ball, ground ball, out, and other. The ``other'' outcomes are those for which the ball is not put into play \cite{null2009}. 
\begin{table}[ht]
\centering
   \begin{tabular}{cccc} 
\hline
Fly Ball & Ground Ball & Out & Other \\
 \hline   
 HR &  HR & Fly Out & Interference \\
 Triple &  Triple & Ground Out & Hit by pitch \\
 Double & Double & Strike Out & Base on balls \\
 Single & Single & & \\
 \hline
 \end{tabular}
 \caption{Outcomes from at-bats as given in \cite{null2009}}\label{tab:components}
 \end{table}
The data consist of outcomes of plate appearances for Major League Baseball players from the years 2000 - 2010. There are 3136 unique players in the data set, and fourteen outcomes for each at bat, as given in the body of Table \ref{tab:components}. Ages of batters range from 20 to 49.
Other variables in the data set include batter ID, batter birth year, handedness of batter (right-handed or left-handed), handedness of pitcher (right-handed or left-handed), and total number of plate appearances. For the analysis in this paper, plate appearance information for batters against only right-handed pitchers (RHP) is used. Both left-handed (LH) and right-handed (RH) batters are included, and they were split into three age groups. The ages represented young batters (age $\leq 26$), experienced batters (age $\geq 35$), middle aged batters ($26 < \text{age} < 35$). These age divisions correspond to the quartiles of age for the entire data set.  
For convenience, we grouped the original 14 outcomes into six components: home runs, singles, doubles, triples, outs, and other. The home runs, singles, doubles, and triples categories include fly ball and ground ball varieties of each. Similarly, the out component includes both ground outs and fly outs. The ``other" component groups the outcomes that do not reflect the batter's skill to hit the ball: catcher interference, hit by pitch, and base on balls.  

The data span 11 seasons; however, not every player plays in every season. For example, 466 players played only one season, 311 players played 2 seasons, and 87 players played all 11 seasons. Because some players played for multiple seasons, the same player can be measured multiple times over the span of the data set. Instead of examining plate appearances for all players in all seasons from 2000-2010 together, we examine results from the 2000 and 2010 seasons, which minimizes the dependence due to players' participation in multiple seasons, and illustrates changes in the composition of the outcomes between seasons. We assume that outcomes for each player within a season are independent of the outcomes of the other players.

\begin{figure*}[htp]
  \centering
  \subfigure[2000 Season]{\includegraphics[scale=0.3]{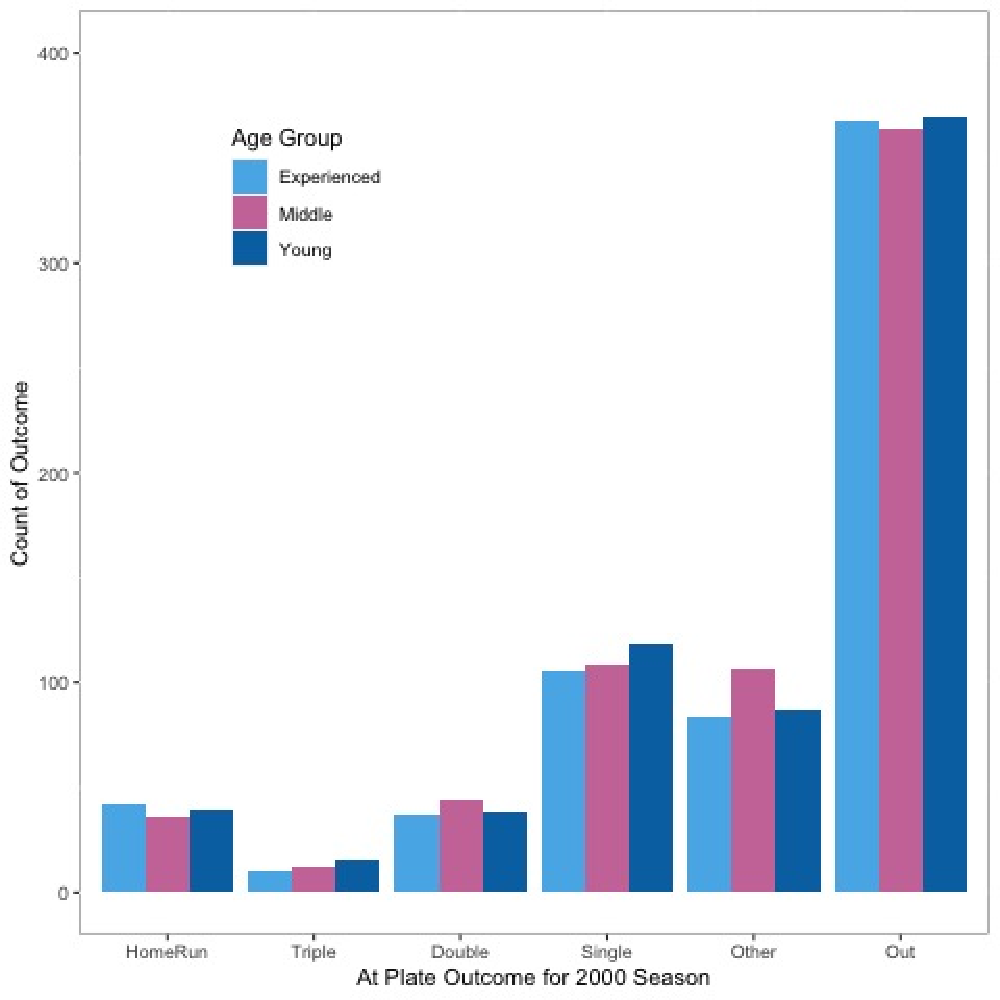}}\quad
  \subfigure[2010 Season]{\includegraphics[scale=0.3]{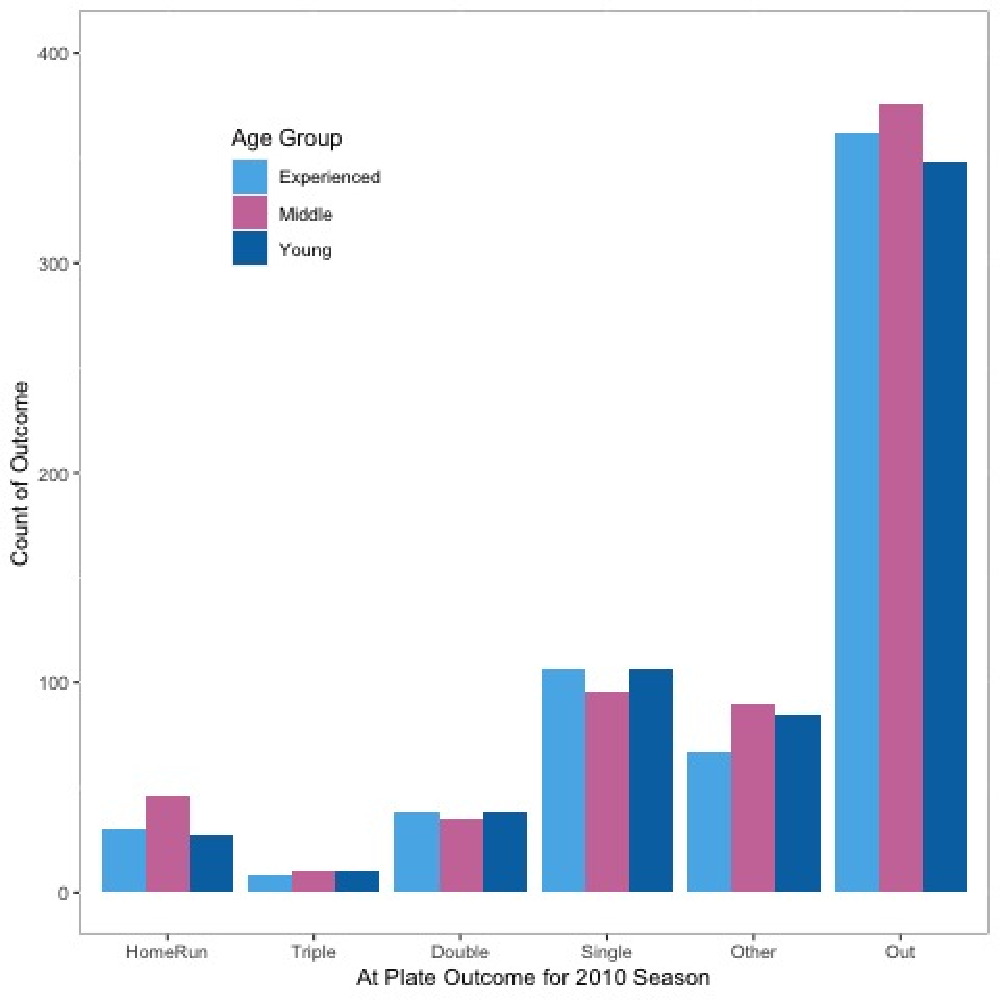}}
   \caption{The number of outcomes for each of six components for experienced (light blue), middle-aged (pink), and young (dark blue) batters. The components are listed on the horizontal axis. The height of each bar represents the frequency of the outcome.}\label{fig:3seasons}
\end{figure*}

Figure \ref{fig:3seasons} shows the number of each outcome by age group for each of the 2000 and 2010 seasons. Each panel corresponds to a different season. The bars are colored light blue for experienced players (the left bar in each cluster of outcomes), magenta for middle aged players (the middle bar in each cluster), and dark blue for the younger players (the right bar in each cluster). Note that a player who is ``young'' in 2000 might be ``experienced'' by 2010; therefore, players who are active for several seasons will change age group membership across time. 

Some aspects of the outcomes are the same regardless of season. For example, outs are always the largest component and triples are always the smallest. This is true regardless of batter age. However, there is some evidence of a difference in the frequency of outcomes by age group for the 2000 and 2010 seasons. We see that in 2000, the number of outs per age group were approximately equal. In 2010, the middle players had the most outs. We can see other changes in frequency of outcomes for the age groups for other components. Using these data, we wish to determine whether changes in the distribution of plate appearance outcomes differ by batter age and, if so, in what manner that distribution differs. 

To facilitate comparison among components, seasons, and age groups, where the absolute number of outcomes differs, the total of the at-bat outcomes for each component within each season can be divided by the total number of plate appearances within that season to obtain a proportion of plate appearance outcomes for a given age group within a season. The total proportion of outcomes within a season equals 1; therefore, the proportions of outcomes must sum to 1. In other words, the data consist of a fixed number of components  where their proportions are dependent due to the fact that their sum is a constant. This is a defining characteristic of compositional data.

More formally, let $\mathbf{X}=(X_1,X_2,...,X_k)$ be a $k$-dimensional non-negative random vector such that $\sum_{j=1}^k X_{j} = 1$.  A compositional data set is a random sample of $n$ observations, $\mathbf{X}_1,\mathbf{X}_2,...,\mathbf{X}_n$, from the joint distribution of $\mathbf{X}$. For the plate appearance data, there are $k=6$ components, each represented by a vector of counts for each batter within each season. Since the counts are all greater than or equal to 0, the data can be represented as proportions or percentages of a whole and is thus a compositional data set. The component's corresponding proportion relative to the batter's total plate appearance counts satisfies the constraint $\sum_{j=1}^k X_{j} = 1$ and thus lives on the unit simplex. Even if the original counts were statistically independent for each component, the variables within the composition are statistically dependent due to the constraint.

\section{ The Nested Dirichlet Distribution}
Compositional data are often treated as if each component is independent, resulting in multiple independent statistical tests, one for each of the $k$ components across the $g$ groups being compared \cite{aitch1982, maugard}. \cite{gloor2017} makes a case against this for microbiome data, which has similar properties to the baseball outcome data in that there are a fixed number of components and the proportion of observations across all components should sum to a constant.

\subsection{Extant Methods for Compositional Data Analysis}
Compositional data has been previously analyzed via transformation of the component proportions with a log ratio, such as the arithmetic log ratio (ALR), centered log ration (CLR), or isometric log ratio (ILR) \cite{aitch1982, vanden2013}. 
Regardless of the log-ratio employed, the procedure for analysis of composition data involves transforming proportions using the chosen LR, and then using the transformed values instead of using the original counts in a classical parametric model. Classical methods based on the Normal distribution, including MANOVA, can be applied to the transformed data because the composition is transformed from the simplex to the entire plane. Once a satisfactory model has been found, its parameter estimates are back-transformed to obtain prediction proportions of individuals within each component. There are several R packages to apply log ratio transforms to compositional data, such as the {\texttt compositions} package \cite{PackageCompositions}. 

Another method of analyzing compositional data is to use a probability distribution whose support lies on a simplex, which acknowledges the inherent structure of the data.  The most common and well known distribution used for compositional data is the Dirichlet Distribution and is given by 
\begin{equation*}
f({\bf x}\vert\alpha)=\frac{1}{B(\alpha)}\prod_{j=1}^k x_j^{\alpha_j-1}\qquad 0\le x_j \le 1; j=1, \ldots , k
\end{equation*}
where the vector ${\bf x}$ is constrained to the unit simplex through $\Sigma_{j=1}^k x_j = 1$. For a composition made up of $k$ variables, the appropriate Dirichlet distribution will have $k$ parameters, $\bf{\alpha}=(\alpha_1,\ldots , \alpha_k )$.  The $\alpha_j$ can be thought of as counts from a prior or a current study. $A = \sum_{j=1}^{k} \alpha_j$ is known as the precision. The mean for each variable $\pi_j := \frac{\alpha_j}{A}$ and the variance is $\sigma_j^2 = \frac{\pi_i(1-\pi_i)}{A+1}$; therefore, components with the same mean must also have the same variance.  The covariance between any two components is non-positive \cite{null2009}.  A strong correlation under the DD does not necessarily correspond to the linear elliptical relationship typically seen in a multivariate setting. 



\begin{figure}[htp]
  \centering
  \subfigure[Correlation for All Players]{\includegraphics[scale=0.3]{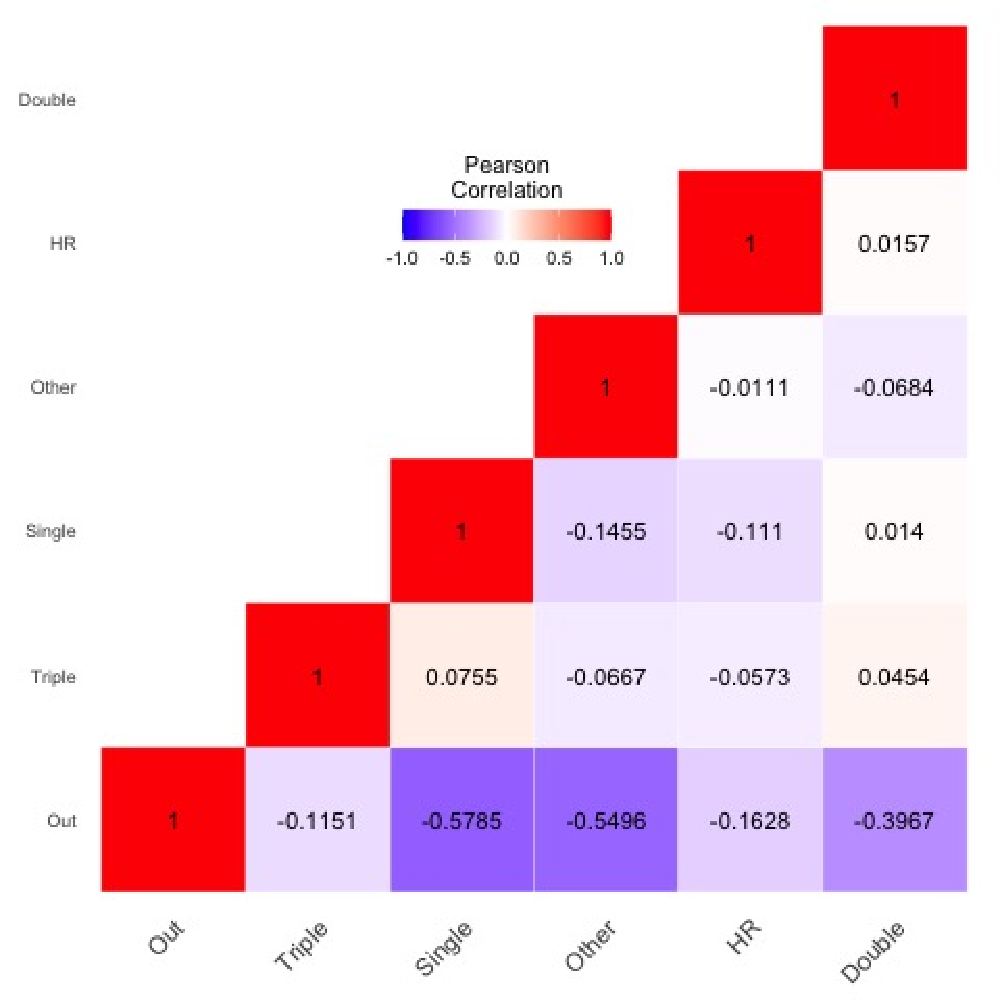}}\quad
  \subfigure[Correlation for Young Players]{\includegraphics[scale=0.3]{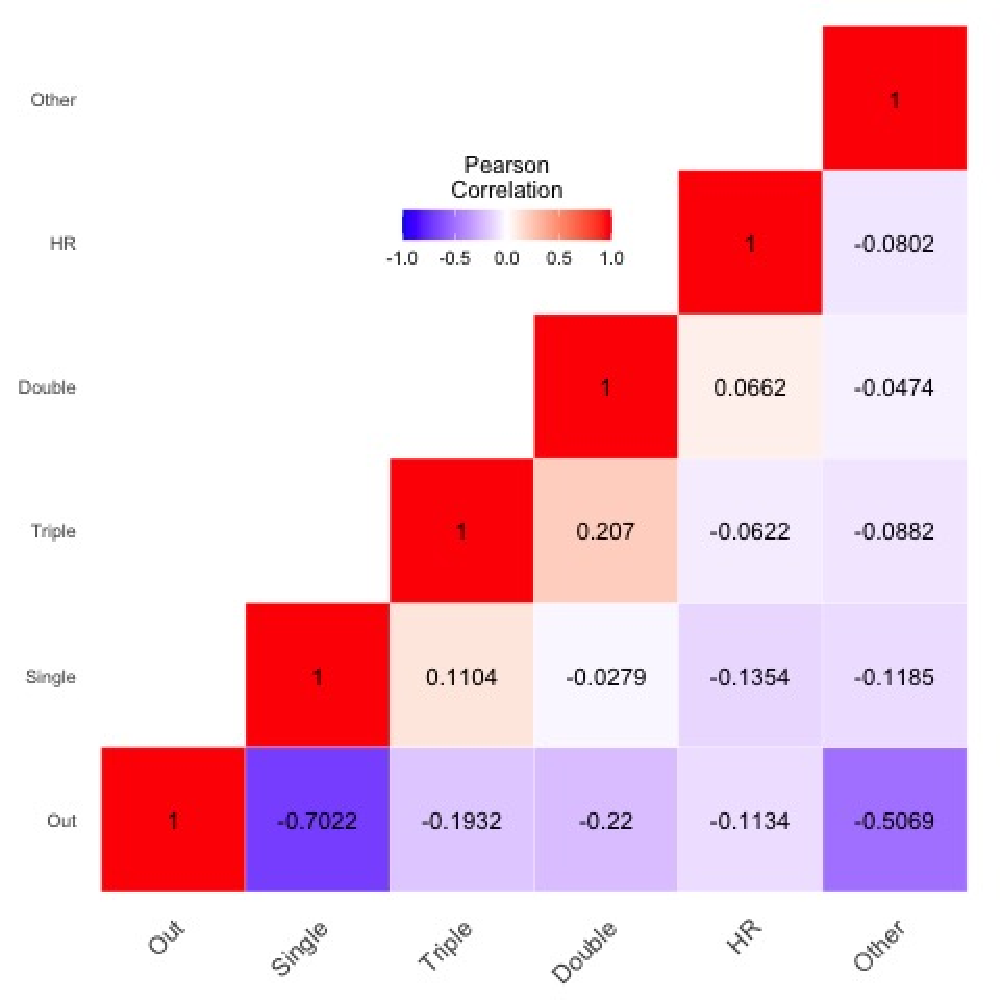}}\quad
  \subfigure[Correlation for Middle Players]{\includegraphics[scale=0.3]{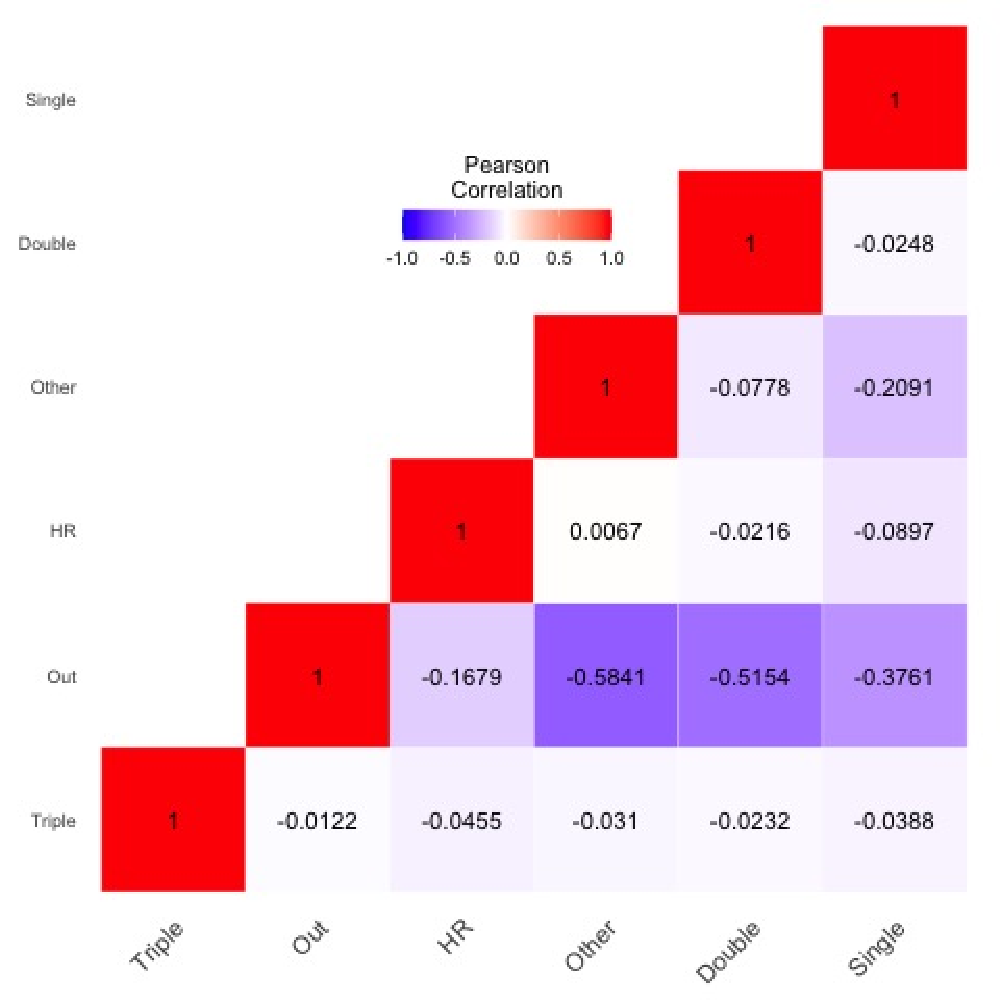}}\quad
  \subfigure[Correlation for Mature Players]{\includegraphics[scale=0.3]{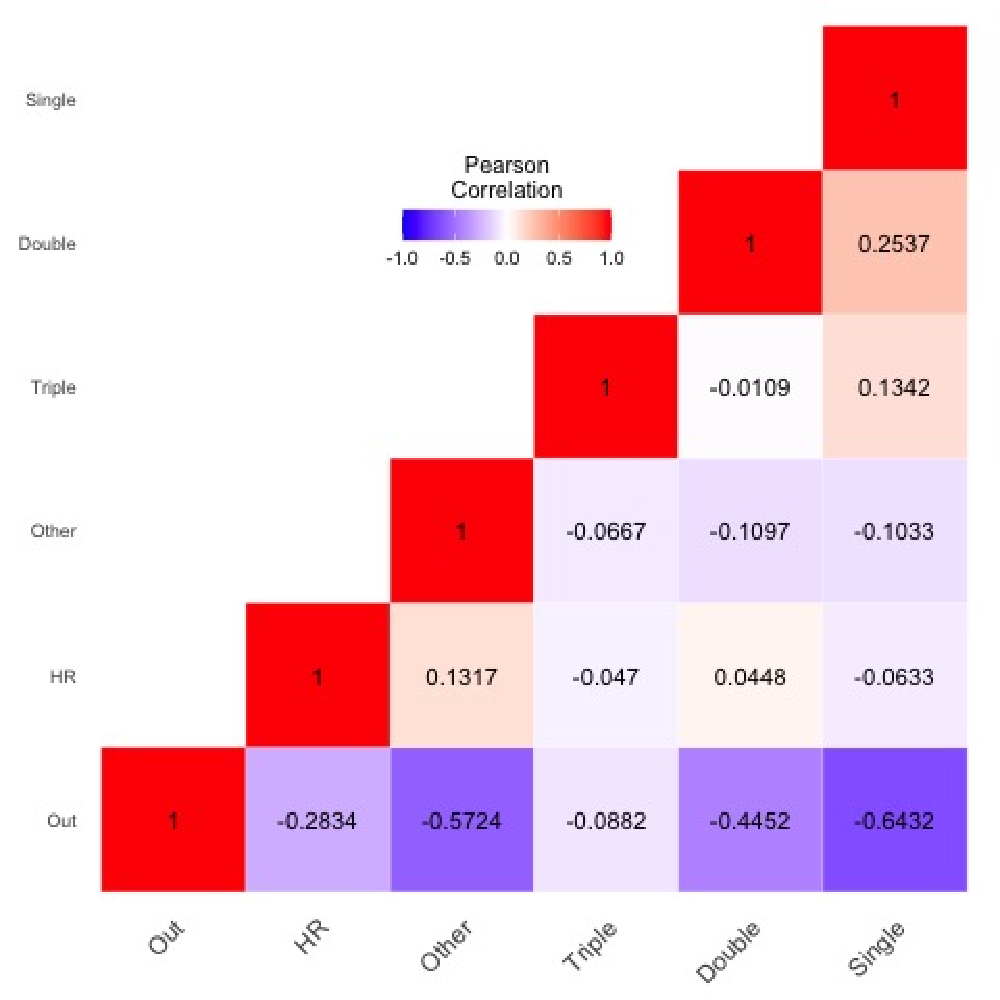}}
  \caption{Pearson's correlation rendered in heatmaps for the 2010 season. (a) Correlation for all players, (b) correlation structure for young players (c) correlation structure for middle-age players and (d) correlation structure for experienced players. }\label{fig:2010cormat}
\end{figure}

Figure \ref{fig:2010cormat}(a) shows a heatmap of the overall correlation structure for  the 2010 season. Figures \ref{fig:2010cormat}(b - d) show correlation heatmaps for the young, middle, and experienced players, respectively. For all correlation matrices, negative correlations are represented by shades of purple. The darker the shade of purple, the larger the correlation in magnitude. Positive correlations are represented by shades of red, where darker shades of red indicate larger magnitude correlations.  The components are along the horizontal and vertical axes. They are ordered by magnitude of correlation rather than by name of the component. For example, if we look at the correlation between singles and triples, the overall correlation is $0.0755$, the correlation for the young players is $0.1174$, the correlation for the middle players is $-0.0388$, and the correlation for experienced players is $0.1342$. Note further that three of these correlations are positive, which violates a property of the Dirichlet distribution. Clearly, we have positive correlations between components that are too large to ignore. Furthermore, we do not want to assume that components with the same correlation will have the same variance. Therefore, we need a more general distribution for comparison of the composition of at-bat outcomes for young, middle-aged, and experienced players.

\subsection{The Nested Dirichlet Distribution} 
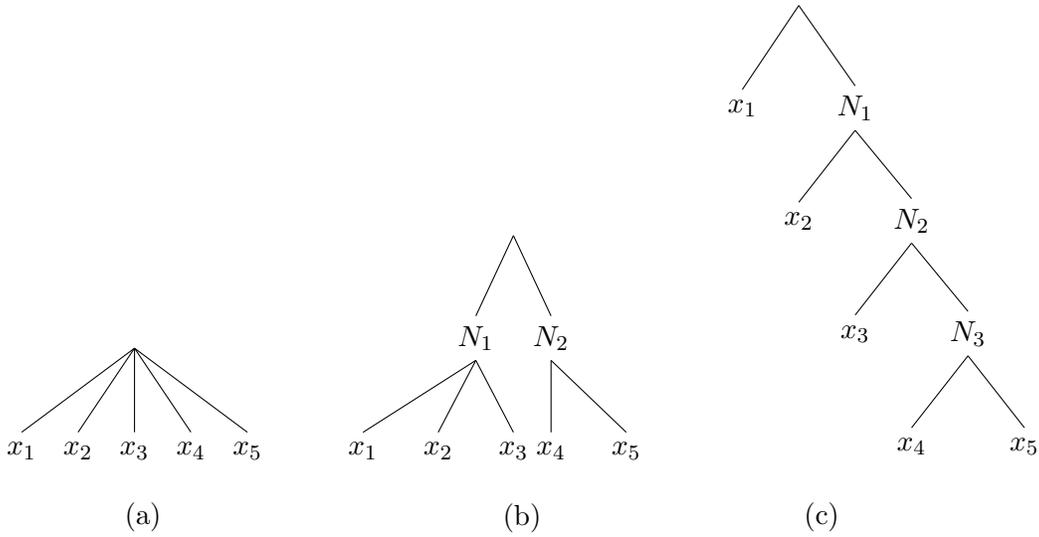
\begin{figure}
\centering
\begin{tikzpicture}
\node{}[sibling distance = .75 cm]
    child { node {$x_1$} }
    child { node {$x_2$} }
   child { node {$x_3$} }
   child {node {$x_4$}}
   child {node {$x_5$}};
    \node at (0.5,-2,1) {(a)};
\end{tikzpicture}
\qquad
\begin{tikzpicture}
\node{}[sibling distance= 1 cm]
    child { node {$N_1$} 
        child { node {$x_1$} }
        child { node {$x_2$} }
        child { node {$x_3$}}
        child[missing]
    }
    child { node {$N_2$}
        child[missing]
        child { node {$x_4$} }
        child { node {$x_5$} }
 };
    \node at (0.5,-3.5,1) {(b)};
\end{tikzpicture}
\qquad
\begin{tikzpicture}
\node{}
   child { node {$x_1$} }
    child { node {$N_1$} 
        child { node {$x_2$} }
        child { node {$N_2$} 
        		child{node{$x_3$}}
        		child{node{$N_3$}
                   child{node{$x_4$}}
        		  child{node{$x_5$}}
            }}};
        \node at (0.5,-6.75,0.5) {(c)};
\end{tikzpicture}
\caption{Comparison of structure for variations on the Dirichlet distribution: (a) standard Dirichlet (b) Nested Dirichlet Distribution (c) Generalized Dirichlet distribution. Each distribution has 5 components. }\label{fig:diffDDs}
\end{figure}

The nested Dirichlet distribution (NDD) relaxes the constraints that variables with the same mean must have the same variances and allows for the covariance between variables to be nonnegative \cite{null2009}. This distribution has also been called the Dirichlet-tree distribution \cite{minka1999, tang2017} and the hyper-Dirichlet type I distribution \cite{dennis1991}.  These distributions are derived by incorporating latent variables which we will refer to as interior nodes. Further, the distributions can be visualized using a tree diagram. Figure \ref{fig:diffDDs} compares tree diagrams of the standard Dirichlet distribution with two nested Dirichlet distributions. Each distribution has $k=5$ components, $\mathbf{x}=(x_1,x_2,x_3,x_4,x_5)$.  For a standard Dirichlet distribution, all components for the Dirichlet are ``nested'' under one node, which is the root node. 
 In addition to the root node, the two NDD trees posses two interior nodes denoted $N_1$ and $N_2$ in which subsets of the components can be nested within. 

Unlike tree structures found in regression and classification models, incorporating interior nodes to generate binary splits within the tree are allowed but not required of the NDD as illustrated in Figure \ref{fig:diffDDs}.  It should be noted, however, that the tree in Figure \ref{fig:diffDDs} (c) has its interior nodes placed sequentially to create a cascade effect of binary splits.  This special case is commonly referred to as the Generalized Dirichlet distribution (GDD) \cite{gdd}. The naming is somewhat confusing given that the NDD generalizes the GDD, but both distributions generalize the DD. The GDD also illustrates the fact that the maximum number of nodes that a tree can have, including the root node, is $k-1$.

\begin{figure}
    \centering
   \begin{tikzpicture}
    \node[circle, draw, minimum size = 3em]{Root}[sibling distance= 5 cm, level distance = 2 cm]
        child {node[circle, draw, minimum size = 3em] {$N_1$} [sibling distance = 2 cm] 
            child{node[circle, draw, minimum size = 3em] {$x_1$}
            edge from parent node[left, xshift=-0.2cm] {\Large $\alpha_1$}}
            child {node[circle, draw, minimum size = 3em] {$x_2$}
            edge from parent node[left, xshift=-0.0cm] {\Large $\alpha_2$}}
            child {node[circle, draw, minimum size = 3em] {$x_3$}
            edge from parent node[right, xshift=0.2cm] {\Large $\alpha_3$}}
            edge from parent node[left, xshift=-0.2cm]{\Large $\alpha_6$}
        }
        child {node[circle, draw, minimum size = 3em] {$N_2$} [sibling distance = 2 cm]
            child {node[circle, draw, minimum size = 3em] {$x_4$}
            edge from parent node[left, xshift=-0.2cm] {\Large $\alpha_4$}}
            child{node[circle, draw, minimum size = 3em] {$x_5$}
            edge from parent node[right, xshift=0.2cm] {\Large $\alpha_5$}}
            edge from parent node[right, xshift=0.2cm]{\Large $\alpha_7$}
        };
    \end{tikzpicture}
    \caption{Tree diagram example for a NDD with corresponding parameters.}
    \label{fig:exampleNDD}
\end{figure}
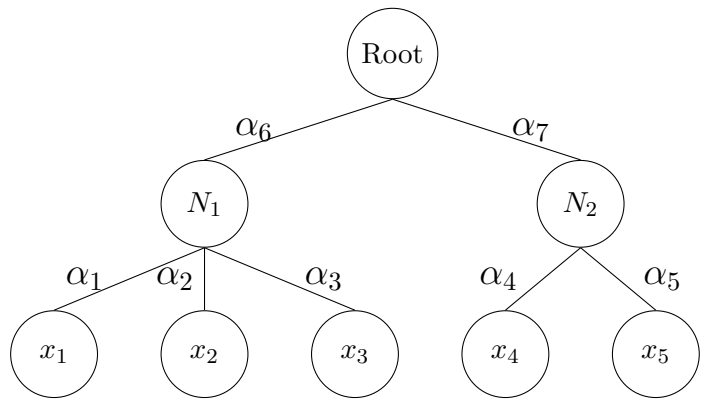

The expression for the NDD for an arbitrary tree is notationally cumbersome \cite{dennis1991}. To help gain insight and to motivate a general testing procedure, we will use an example for a specific tree. Figure \ref{fig:exampleNDD} provides a tree visualization of a NDD with $k=5$ components.  Additional notation along the edges of the tree denotes the parameters of the distribution. The joint density of the compositional vector $\boldsymbol{X}$ that follows a NDD given the tree in Figure \ref{fig:exampleNDD} is proportional to Equation \ref{eq:ndd}, up to a normalizing constant.
\begin{equation}
  f(\bm{x}|\bm{\alpha}) \propto  \prod_{j=1}^k x_j^{\alpha_j-1} \left( \sum_{j=1}^3 x_j \right)^{\alpha_6-\sum_{j=1}^3 \alpha_j} \left( \sum_{j=4}^5 x_j \right)^{\alpha_7-\sum_{j=4}^5 \alpha_j} \quad  0\leq x_j \leq 1\ \text{for}\ j=1, \dots ,k.
  \label{eq:ndd}
\end{equation}
In Equation \ref{eq:ndd},  $\boldsymbol{\alpha}$ is the vector of parameters $(\alpha_1,\alpha_2,...,\alpha_7)$.  The number of parameters in a NDD will be equal to the sum of the number of components $k$ and the number of internal nodes, or the number of edges within the tree.  It is easily verified that if $\alpha_6=\sum_{j=1}^3 \alpha_j$ and $\alpha_7=\sum_{j=4}^5 \alpha_j$, then the joint density reduces to the standard Dirichlet density \cite{dennis1991}. From the tree diagram perspective, if the parameter associated with an internal node is equal to the sum of immediate children's parameters, then the node is removed from the tree.

The normalizing constant for Equation \eqref{eq:ndd} is the reciprocal of the product of three beta functions $B(\alpha_1,\alpha_2,\alpha_3)$, $B(\alpha_4,\alpha_5)$, and $B(\alpha_6,\alpha_7)$, and provides some insight as to how the distribution was first derived \cite{dennis1991}.  It is also worth noting that the parameters within each beta function correspond to one ``layer" of the tree as depicted in Figure \ref{fig:exampleNDD}. Since the three beta functions correspond to normalizing constants for 3 separate Dirichlet densities, it seems reasonable to assert that the NDD density function is related to 3 separate DD functions.  This is indeed the case. 

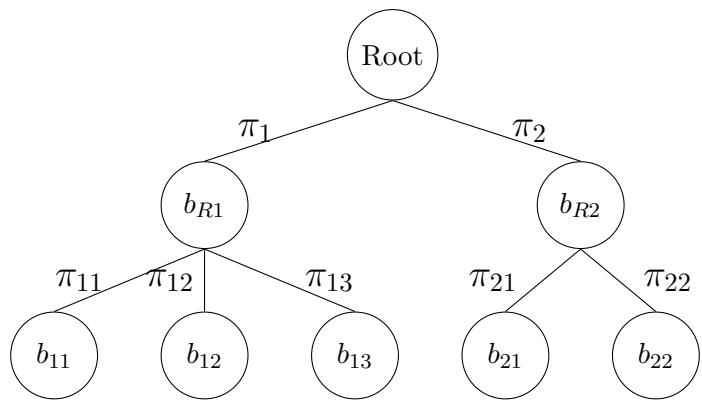
\begin{figure}
    \centering
   \begin{tikzpicture}
    \node[circle, draw, minimum size = 3em]{Root}[sibling distance= 5 cm, level distance = 2 cm]
        child {node[circle, draw, minimum size = 3em] {$b_{R1}$} [sibling distance = 2 cm] 
            child{node[circle, draw, minimum size = 3em] {$b_{11}$}
            edge from parent node[left, xshift=-0.2cm] {\Large $\pi_{11}$}}
            child {node[circle, draw, minimum size = 3em] {$b_{12}$}
            edge from parent node[left, xshift=-0.0cm] {\Large $\pi_{12}$}}
            child {node[circle, draw, minimum size = 3em] {$b_{13}$}
            edge from parent node[right, xshift=0.2cm] {\Large $\pi_{13}$}}
            edge from parent node[left, xshift=-0.2cm]{\Large $\pi_1$}
        }
        child {node[circle, draw, minimum size = 3em] {$b_{R2}$} [sibling distance = 2 cm]
            child {node[circle, draw, minimum size = 3em] {$b_{21}$}
            edge from parent node[left, xshift=-0.2cm] {\Large $\pi_{21}$}}
            child{node[circle, draw, minimum size = 3em] {$b_{22}$}
            edge from parent node[right, xshift=0.2cm] {\Large $\pi_{22}$}}
            edge from parent node[right, xshift=0.2cm]{\Large $\pi_2$}
        };
    \end{tikzpicture}
    \caption{Mean vectors of subcompositions transformed from a NDD.}
    \label{fig:exampleNDDmeans}
\end{figure}

Letting $N_1=x_1+x_2+x_3$ and $N_2=x_4+x_5$, we transform the original compositional vector $\bf x$ to branch proportions.   That is, we express the original variables as smaller dimensional compositions (subcompositions) relative to each of the internal nodes of the tree. The transformation to subcompositions is illustrated in Figure \ref{fig:exampleNDDmeans}. The first subcomposition under the root node is defined as $\mathbf{b_{R}}=(b_{R1}=N_1,b_{R2}=N_2)$ respectively. The subcomposition under the first internal node is defined as $\mathbf{b_{1}}=(b_{11}=x_1/N_1,b_{21}=x_{2}/N_1,b_{13}=x_3/N_1)$.  The final subcomposition under the second internal node is defined as $ \mathbf{b_{2}}=(b_{21}=x_{4}/N_2,b_{22}=x_{5}/N_2$).  The transformation from the original composition to the subcompositions is a one-to-one transformation and it can be shown that each subcomposition is independent and follows standard Dirichlet distributions \cite{dennis1991}.  For our example, it follows that $\mathbf{b_R}\sim DD(\alpha_6,\alpha_7)$, $\mathbf{b_1}\sim DD(\alpha_1,\alpha_2,\alpha_3)$, and $\mathbf{b_2}\sim DD(\alpha_4,\alpha_5)$. 

As stated previously, the NDD relaxes the constraints of the distributional characteristics of the standard DD by allowing for unique variances for components with the same mean as well as allowing for positive correlations.  The mean, variances, and covariances of the NDD can be computed by deriving the first and second moments using the independence of the subcompositions and the moments of the standard DD \cite{dennis1991}.  For example, denote the mean vector of the NDD defined in Equation \eqref{eq:ndd} as $\mathbf{p}=(p_1,p_2,p_3,p_4,p_5)$ and let the mean vector of the subcompositions be denoted as $\mathbf{\pi_R}=(\pi_{R1},\pi_{R2})$, $\mathbf{\pi_1}=(\pi_{11},\pi_{12},\pi_{13})$, and $\mathbf{\pi_2}=(\pi_{21},\pi_{22})$. The mean parameters of the subcompositions are displayed in Figure \ref{fig:exampleNDDmeans}.  Since $x_1$ can be expressed in terms of the subcompositions' components, the expectation  $p_1=E(x_1)=E(b_{R1} b_{11})=E(b_{R1})E(b_{11})=\pi_1 \pi_{11}$ due to the independence of at each layer. In general, the mean of a component, $x_i$, from the NDD is the product of the mean of the branch proportions along the path starting at the root node down to the terminal node associated with $x_i$. The NDD further generalizes the DD in that variables nested under unique nodes can have the same mean but different variances \cite{null2009}.

\section{Overall Test for G=3 groups} \label{sec:LRT}
A likelihood ratio test for the null hypothesis $H_0:\mathbf{p}_1 = \mathbf{p}_2$, where $\mathbf{p}_g$, $g = 1, 2$, represent mean vectors from a DD or a NDD has been previously derived \cite{tml2025}. Here, we extend the previous results to $G=3$ groups. The extension to $G > 3$ groups is straight-forward. 

Suppose that observations from $G=3$ distinct groups are obtained from the same NDD tree structure as depicted in Figure \ref{fig:exampleNDD}; however, the mean vectors are potentially different, and denoted as $\mathbf{p_1}$, $\mathbf{p_2}$, and $\mathbf{p_3}$.  We wish to test the hypotheses
\begin{center}
$H_0: \mathbf{p_1} = \mathbf{p_2} = \mathbf{p_3}$ \linebreak
 $H_A$: at least one pair of mean vectors differs.
 \end{center} 
It should be noted that under the mean parameterization of the NDD with $l$ interior nodes, there are $3 \times l$ nuisance parameters. These parameters correspond to the precision parameters of each of the $l$ DD layers of the NDD under the subcomposition transformation framework and across the three groups. Similar to the mean vectors, we denote these parameter vectors of length $l$ for each group as $\mathbf{A_1}$, $\mathbf{A_2}$,$\mathbf{A_3}$.

A likelihood ratio test (LRT) statistic can be derived by utilizing the fact that changes in the mean vector of the NDD across groups correspond to at least one difference in the mean vectors of the DD layers under the subcomposition transformation. Letting $\bm{\pi_{l,g}}$ denote the mean vector of the $l^{th}$ subcomposition for group $g$, an equivalent set of hypotheses can be defined as 
\begin{equation}
\label{eqn:nddtest}
\begin{split}
\text{H}_0: & \hspace{0.1in} \bm{\pi_{l,1}} = \bm{\pi_{l,2}}= \bm{\pi_{l,3}} \hspace{0.1in} \text{For all $l$ subcompositions}\\
\text{H}_1: &  \text{ At least one pair of mean subcomposition vectors differ.}
\end{split}  
\end{equation}

Upon transformation of the original data to subcompositions with the number of components at each layer denoted $K_l$, a likelihood ratio statistic can be created to compare the mean vectors of each layer using the standard DD likelihood, denoted $\Lambda_l$, on the $-2log$ scale.  To obtain MLEs for the mean and precision parameters under the unrestricted and restricted cases, we employ a two step maximization procedure \cite{tml2025,luedeker22}.  In both the null and alternative cases, first we treat the mean as fixed and maximize over the precision parameters. Then, we maximize over the mean vectors while holding the precision vectors fixed.  The updates continue iteratively until a convergence criterion is reached. The degrees of freedom for $\Lambda_l$ is $GK_l-(K_l-1+G)$ where $GK_l$ and $K_l-1+G$ are the number of parameters in the unrestricted and restricted cases respectively.

Alternatively, one can obtain $\Lambda_l$ by using Dirichlet regression with a single categorical covariate \cite{hijazi2009modeling}.  Due to the independence of each subcomposition, the overall LRT statistic for the entire NDD tree when comparing $G$ groups is

\begin{equation}
\label{eqn:lrt}
\Lambda_{overall} = \sum_{l=1}^{L} \Lambda_l \sim \chi^2_{v}.
\end{equation}
where the degrees of freedom $v$ is the sum of the degrees of freedom for each $\Lambda_l$, $v=\sum_{l=1}^{L} GK_{l}-G-K_{l}+1$ due to the independence of the subcompositions.

\section{Nested Dirichlet Methods Applied to Plate Appearance Outcomes \label{sec:application}}

In order to use the NDD to analyze the baseball outcome data, we need to define a tree that specifies the nesting structure of the components. The nesting structure  in certain data sets is naturally set according to biological or other scientific relationships. For example, microbes are classified to accepted categories of Domain, Kingdom, Phylum, etc \cite{cain2023}. In other words, taxa to which microbes belong are represented by a scientifically defined tree. In metagenomic classification of microbial experiments, the tree structure is given, and the data can be used to assign proportions of microbes at each layer of the tree, with the constraint that child proportions  sum to 1 conditioned on their parent component \cite{gioia2021estimation, gloor2017}.

\begin{figure}
\begin{minipage}[c]{0.4\linewidth}
\includegraphics[width=\linewidth]{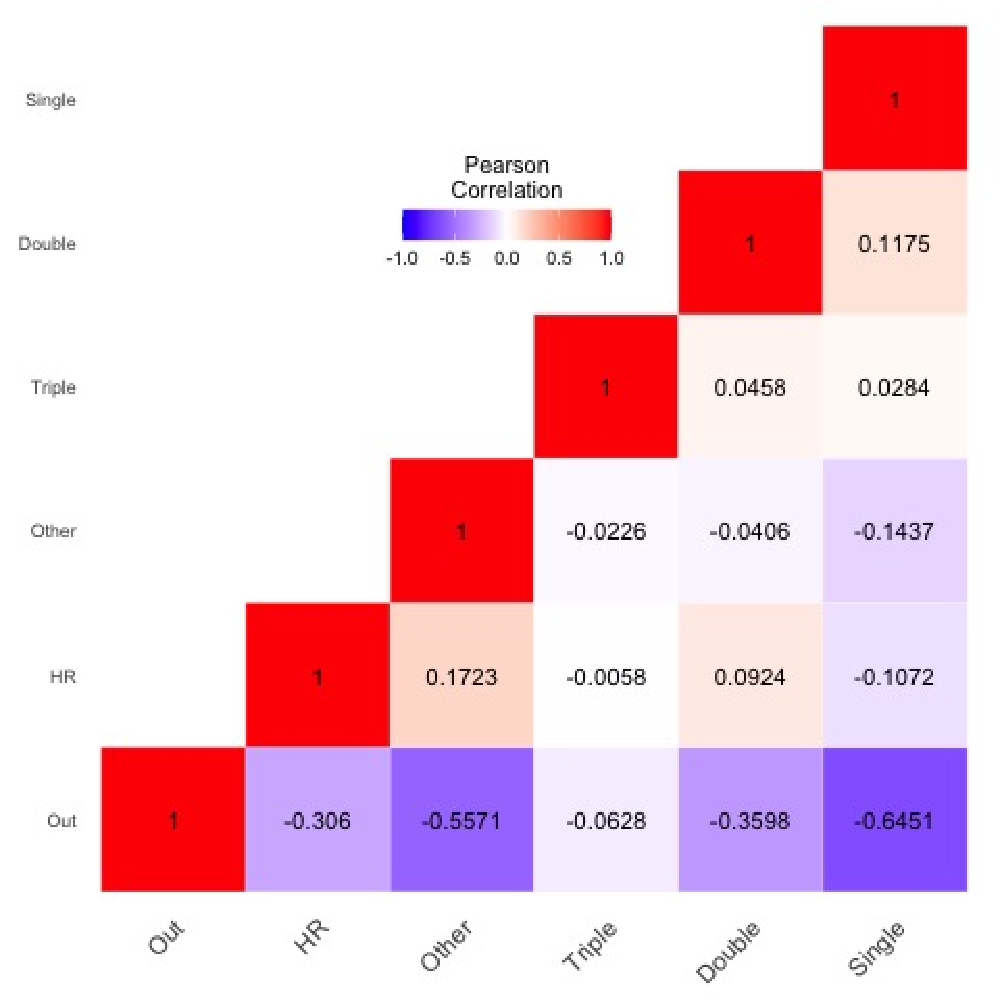}
\caption{Correlation Heatmap for 2000 Season}
\label{fig:cormap2000}
\end{minipage}
\hfill
\begin{minipage}[c]{0.4\linewidth}
\includegraphics[width=\linewidth]{overallHmap2010.eps}
\caption{Correlation Heatmap for 2010 Season}
\label{fig:cormap2010}
\end{minipage}%
\end{figure}

For the baseball outcome data, no scientifically defined tree exists. The groupings of the outcomes into six components are reasonable; however, the components themselves have no intuitive structure. It is the correlation structure among the components that defines the tree. Figures \ref{fig:cormap2000} and \ref{fig:cormap2010} show heatmaps of the Pearson correlation among the six groups of outcomes for the 2000 and 2010 seasons. Red color is for positive correlation and blue is for negative correlation. The deeper the saturation of the color, the greater the magnitude of the correlation. White color is for correlations close to zero. We see that there are some saturated red squares, which indicate that there are positive correlations among the outcomes. In particular, ``Other'' outcomes are positively correlated with home runs for both the 2000 and 2010 seasons. There is also a positive correlation between singles and doubles for the 2000 season that is close to 0 for the 2010 season. The differences in correlation structure for the 2000 season and the 2010 indicate that different tree structures for the two seasons are appropriate.

To find a tree that reflects the correlation structure in the data, we use an algorithm presented in \cite{luedeker22}. The algorithm utilizes the fact that any NDD tree can be generalized by adding additional interior nodes such that every split in the tree is a binary split \cite{minka1999, null2008}.  For this discussion we will denote the likelihood of the NDD as $L(\mathbf{\alpha}|\mathbf{x})$. Steps of the algorithm are given below. More details on the tree-finding algorithm, as well as the effects of overspecification and a methods of pruning, are the subject of a forthcoming manuscript. 
\begin{algorithm}
\caption{Tree-Finding Algorithm}\label{alg:tree}
\begin{algorithmic}[1]
\State Fit all variables to the standard Dirichlet distribution. Obtain $L(\hat{\alpha}\vert \mathbf{x})$.
\State Obtain $L(\hat{\mathbf{\alpha}}\vert \mathbf{x})$ for each of the possible permutations of variables for the two NDDs with binary splits.
\State Make a decision to incorporate interior nodes based on the smallest $-2L(\hat{\mathbf{\alpha}}\vert \mathbf{x})$ value or a penalized version ({\it e.g.} AIC or BIC).
\State For every split that contains more than 2 variables, convert to branch proportions (to ensure that proportions sum to 1) and apply steps 1-3 again.
\State Continue until splitting is no longer favored  or when the tree consists entirely of binary splits.
\end{algorithmic}
\end{algorithm}

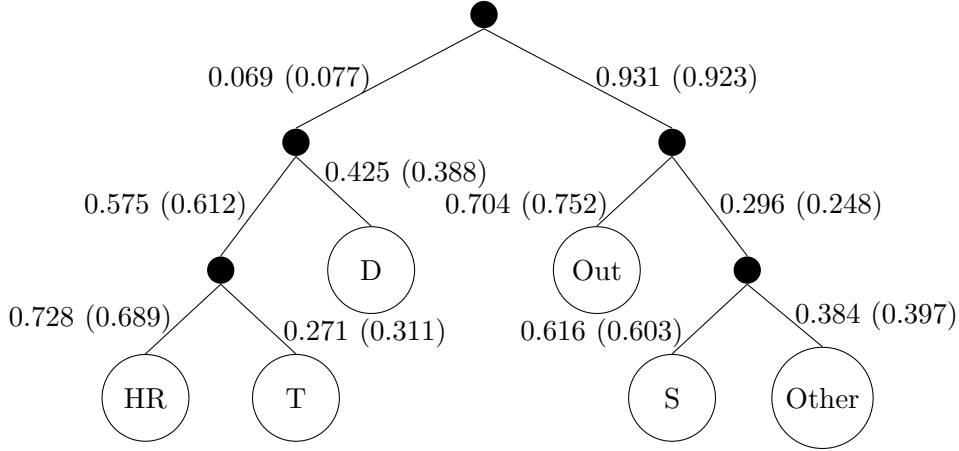
\begin{figure}
	\leavevmode\hidewidth\begin{tikzpicture}
		\node[fill, circle] {}[sibling distance = 5 cm, level distance = 1.7cm] 
		child{node[fill, circle] {} [sibling distance = 2cm] 
			child{node[fill, circle] {} [sibling distance = 2cm] 
				child{node[circle, draw, minimum size = 3em]{HR} 
					edge from parent node[left]{0.728 (0.689)}} 
				child{node[circle, draw, minimum size=3em]{T} 
					edge from parent node[pos=0.7, right]{0.271 (0.311)}} 
				edge from parent node[left]{0.575 (0.612)}    
			}
			child{node[circle, draw, minimum size = 3em] {D} 
				edge from parent node[pos=0.25, right]{ 0.425 (0.388)}}
			edge from parent node[left, xshift=-0.1cm]{0.069 (0.077)} 
		}
		child{node[fill, circle] {} [sibling distance = 2 cm]
			child{node[circle, draw, minimum size = 3em]{Out}
				edge from parent node[pos=0.7, left]{0.704 (0.752)}}
			child{node[fill, circle]{} [sibling distance = 2cm]
				child{node[circle, draw, minimum size = 3em] {S}
					edge from parent node[pos=0.7, left] {0.616 (0.603)}}
				child{node[circle, draw, minimum size = 3em] {Other}
					edge from parent node[right]{0.384 (0.397)}}
				edge from parent node[right]{0.296 (0.248)}
			}
			edge from parent node[right, xshift=0.1cm]{0.931 (0.923)} 
			};
	\end{tikzpicture}\hidewidth\null
    \caption{Tree for the baseball data. The numbers along each branch indicate the percentage of outcomes for that branch. Percentages for the 2000 season are listed first. The percentages for the 2010 season are listed in parentheses.}
    \label{fig:2000tree}
\end{figure}

The overall tree, regardless of age group, using Algorithm \ref{alg:tree}, is given in Figure \ref{fig:2000tree}.  Figure \ref{fig:2000tree} has the components HR, T, and D grouped together on the left-hand side of the tree vs. the outcomes Out, Single, and Other on the right-hand side. The numbers along each branch indicate the percentage of the outcomes \textcolor{blue}{for that branch relative to the immediate parent component}. The percentage for the 2000 season is listed first with the percentage for the 2010 seasons listed in parenthesis. For example, at the root level, approximately 7\% of outcomes are home runs, triples, and doubles while the other 93\% are outs, singles, and others. At the next level of the tree on the left hand side, doubles occur 42\% \textcolor{blue}{of the time relative to the total number of HR, T, and D in the 2000 season and 39\% of the time, again relative to HR, T, and D, in the 2010 season.} The outcomes home run and triple occur the other 58\% (61\%) of the time \textcolor{blue}{relative to HR, T, and D}. The percentages for each level of the tree sum to 100\%. Note that the outcomes that tend to lead to runs are grouped together on the left-hand side, while the outcomes that are less likely to lead to runs are grouped on the right-hand side.  This tree structure is the same for all three age groups within a given season, although the branch proportions might be different for the age groups - in fact our test results show that this is the case.

Utilizing the LRT statistic developed in Section \ref{sec:LRT}, we test the hypothesis that 
\begin{center} 
$H_0: \bm{p_Y} = \bm{p_M} = \bm{p_E}$ \linebreak 
$ H_A:$ at least one pair of mean vectors differs.
\end{center}
\noindent where $\bm{p}$ corresponds to the mean vector of proportions for the specified NDD, and the indices Y, M, and E denote the young, middle-aged, and experienced groups, respectively. 
\begin{table}[ht]
\begin{center}
\begin{tabular}{rcc}
\hline
Level of Tree & LR for 2000 &  LR for 2010 \\
\hline\hline
\textcolor{red}{Total} & \textcolor{red}{51.21 ($ 2.9 \times 10^{-6}$)} & \textcolor{red}{40.86 (0.00012)}\\
\hline\hline
Root & 3.560 & \textcolor{blue}{17.83}\\
\hline
HR$+$T, D & 0.129 & 4.587 \\
\hline
HR \& T & \textcolor{blue}{19.288} & 6.470\\
\hline
Out, S$+$Other & \textcolor{blue}{20.781}& 3.181 \\
\hline
S \& Other &7.416& 8.807\\
\hline
\label{tab:2000tree}
\end{tabular}
\caption{Likelihood ratio statistics (LR) for the 2000 and 2010 seasons. The overall LR is given in the top row, and the LR for each split of the tree is given in each row. The split is labeled in the far left column. Values highlighted in blue font are the largest LR statistics; thus, they are the ones contributing the most to the overall LR.}\label{tab:lrstats}
\end{center}
\end{table}
Results of Overall Test for 2000 \& 2010 Seasons are given in Table \ref{tab:lrstats}. The overall test statistic, $\Lambda_{overall}$, using Equation \ref{eqn:lrt}, is given in red on the first row of the table with its associated p-value in parentheses. There is strong evidence against the null hypothesis that the compositional means among the age groups are equivalent (LR$=51.21$ with $p < 0.0001$ for the 2000 season, and LR$=40.86$ with $p < 0.0001$ for the 2010 season). To gain further insight into which components contribute to the overall LRT, the rest of the table shows the breakdown of the summands for the overall test statistic for each subcompositional layer of the tree. We see that, for the 2000 season, the components with the largest contribution to the overall test statistic are `HR\&T' (home runs and triples) and `Out, S, \& Other' (outs, singles, and other outcomes) on the right branch of the tree.  Looking at \ref{fig:2000tree}, we see that HR, T, and D are aggregated together on the left side of the tree. However, according to \ref{tab:2000tree}, the LR for doubles is $0.129$ and the LR for the combination of home runs and triples is $19.288$. The small LR for doubles indicates that the difference in composition is not due to doubles but rather to triples and home runs. Likewise, on the right hand side of \ref{fig:2000tree}, the LR for Out, S $+$ Other $= 20.781$, while the LR for S\& Other is 7.416. This indicates that the large LR for the right hand side of the tree is due more to outs than it is to singles and other events. For the 2010 season, the component with the largest contribution is the root layer which corresponds to a difference in the mean total proportion of home runs, triples, and doubles, between the three age groups.

Recall that the alternative hypothesis for the LRT in Equation \ref{eqn:lrt} is vague. If the null is rejected, it is not clear from the alternative which two age groups contributed the most to the difference, or even if the difference is among all three age groups. In the ANOVA context, where there is a similarly vague alternative hypothesis, pairwise \textit{post hoc} tests are employed to answer the question of which population means differ between pairs of populations. We employ a similar technique to determine whether the means of the components differ between pairs of populations. 
\begin{table}[ht]
\resizebox{\textwidth}{!}{%
\begin{tabular}{|l|cc|cc|cc|}
  \hline
 & \multicolumn{2}{|c|}{Young vs. Middle} & \multicolumn{2}{|c|}{Young vs. Experienced} & \multicolumn{2}{|c|}{Middle vs. Experienced} \\ 
\hline 
Split & 2000 & 2010 & 2000 & 2010 & 2000 & 2010\\
  \hline
Root & 0.69 & \textcolor{magenta}{17.12}& 3.09 & 0.28 & 1.16 & \textcolor{magenta}{20.97}  \\ 
  HR $+$ T, D & 0.34 & 5.49 & 0.68 & 7.58 & 0.90 & 1.03 \\ 
  HR \& T & \textcolor{magenta}{18.36} & 1.61 & 1.10 & 2.95 & \textcolor{magenta}{26.76} & 1.51 \\ 
  Out, S $+$ Other & 2.83 & 2.54& \textcolor{magenta}{18.15} & 5.33& 8.79 & 5.21\\ 
  S \& Other & 0.65 & \textcolor{magenta}{11.53} & 4.50 & \textcolor{magenta}{13.56} & 7.04 & 2.86 \\ 
\hline
  Total & 22.87 & 38.39 &  27.52 & 29.70& 44.65 &31.58\\ 
  P-value & 0.0004 &$3.3 \times 10^{-7}$ &$4.5 \times 10^{-5}$ &$ 1.7 \times 10^{-5}$ & $ 1.7 \times 10^{-8}$ & $ 7.2 \times 10^{-6}$\\ 
   \hline
   \end{tabular}}
\caption{Numbers in columns are likelihood ratios for pairwise comparisons of $H_0: \bm{p_i} = \bm{p_j}$  vs. $H_A: \bm{p_i} \ne \bm{p_j}$, where $i, j = 1, 2, 3$ and $i \ne j$.}
\label{tab:pairwise}
\end{table}

The results from pairwise tests for 2000 \& 2010 seasons are given in Table \ref{tab:pairwise}, with the comparison of young batters vs. middle-aged batters given in the left most column. We see that the largest value of the likelihood is associated with home runs and triples for the 2000 season. This indicates evidence that the proportions of home runs and triples for younger players are different from those of middle-aged players. We see that this is also true for middle-aged versus experienced players in the right hand column. The analysis suggests that middle-aged players tend to hit more home runs and triples than do young players. Practically speaking, middle-aged players are strong enough to launch the ball into the outfield, but also patient enough to wait for ``their pitch''. Therefore, middle-aged and experienced players are still valuable to the team in terms of run production.
\begin{table}[ht]
\centering
\begin{tabular}{lccc}
\hline
Outcome & Young & Middle & Experienced \\
\hline
Hit by Pitch 2000 &0.06314\%& 0.09916\%& 0.06504\% \\
Hit by Pitch 2010 & 0.04513\%& 0.09594\% & 0.06788\%\\
Intentional Walk 2000 & 0.00076\% & 0.00152\% & 0.00076\%  \\
Intentional Walk 2010 & 0.00038\%& 0.00246\%  & 0.00133\%\\
Base on balls 2000 & 0.59328\% & 0.93685\% & 0.85968\% \\
Base on balls 2010 & 0.47174\% & 0.85247\%& 0.60068\% \\
Outs 2000 & 4.61201\%& 6.73580\%& 5.65997\%\\
Outs 2010 & 4.10443\% & 7.29381\% & 4.88618\% \\
\hline
    \end{tabular}
    \caption{Proportion of walks, outs, and other outcomes for young, middle--aged, and experience players in the 2000 and 2010 seasons.}\label{tab:walks}
\end{table}

For young versus middle-aged players in the 2010 season, the largest likelihoods are at the root and for single vs.~other. Similarly, the largest likelihood is associated with the root in the 2010 season for the pairwise test of middle-aged versus experienced players. For young versus experienced players in 2010, the largest likelihood is for the proportions associated with singles and others. 

Table \ref{tab:walks} shows the proportion of walks, outs, and other outcomes from the data for the 2000 and 2010 season. These descriptive statistics are commensurate with the model, indicating that experienced players tend to walk more than do younger players (.47\% to .60\%, respectively). In 2010, experienced players were more likely to get hit by a pitch than young players, and they were also more likely to be walked intentionally. Getting on base by any method is important to scoring runs. More experienced players seem to able to wait for a good pitch before hitting, which indicates more patience, or perhaps more humility, than younger players. More experienced players contribute by providing fodder for RBIs.

In conclusion, for the 2000 \& 2010 data, we see that there is strong evidence for a difference between all pairs of players based on age. While the tree structures are similar for the 2000 and 2010 seasons, the seasons differ in which components carry the most ``weight''. This is also true for the age groups when we examine pairwise tests. Layered Dirichlet Modeling (LDM) allows the analyst to learn what outcomes, or groups of outcomes, show the largest differences among age groups. We showed that different seasons have different compositions of outcomes, and that there is evidence that young players perform differently than either middle-aged or experienced players. Our analysis shows that more experienced players contribute in tangible ways to run production, which ultimately helps teams to win games.

\section{Conclusion \& Future Work}

While it is clear that athletic ability diminishes with age, it is also known that older MLB players are paid more than younger players. The justification for the higher pay for older players is the value placed on intangible factors such as their experience, perspective, and leadership. While all these are likely true, we use Layered Dirichlet Modeling show that different age groups of players contribute tangibly to run production. LDM allows direct comparison the probabilities associated with six different at-bat outcomes among young (age $\leq 25$), middle-aged ($25 <$ age $< 35$), and experienced (age $\geq 35$) MLB players.

Interest is in differences in the vector of mean proportions from the distribution of outcomes from plate appearances for the three different age groups of batters from 2000 and 2010.
We show that outcomes from plate appearances in baseball are compositional, which means they are statistically dependent. The Dirichlet distribution is a common parametric model for compositional data, but its restrictions on the equality of means and variances and the nonnegativity of covariances between components render it inappropriate for the at-bat outcome data. A generalization of the Dirichlet distribution, the nested Dirichlet distribution (NDD), takes this dependence among components into account while relaxing restrictions on the covariance structure and mean-variance relationship. However, the NDD requires a tree structure to show relationships among outcomes. We used a previously-developed tree finding algorithm to show relationships among at-bat outcomes. Using this tree structure, we developed a LRT to examine differences in the composition of $G\ge 2$ groups when the data come from a NDD. 

Our methodology requires that the tree structure be the same for each of the $G$ groups. Therefore, we calculate a tree using the correlation structure for the pooled data set, regardless of age group. Once the overall tree is fit, we use the data from each age group to find the values of the branch proportions. However, it is possible that one of the age groups is better fit by a different tree. One path for further analysis is to examine the robustness the LRT to the common tree assumption. If the LRT is not robust, then an extension of this method that allows for different tree structures among the groups would give the model more flexibility. 

In this study, we did not have data on defensive outcomes. Clearly, this is another area of interest. We restricted our study to outcomes against right-handed pitchers; however, it would be of interest to see whether the same proportions change when the pitcher is left-handed, or to do a comparison of the composition of outcomes for right-handed vs.~left-handed pitchers. Television and radio commentators often mention how the handedness of the pitcher affects batters with different dominant hands. Therefore, it would be interesting to consider an extension of the $2\times 2$ factorial design in the LDM setting, where one could compare the composition of outcomes for left-handed and right-handed batters when considering a factor for the handedness of the pitcher. Finally,  LDM can be used to examine the effect of rule changes, such as raising the pitcher's mound or the addition of the pitch clock, have on at-bat outcomes. 

There are many applications and extensions for this methodology. For example, metagenomic and flow cytometry data are compositional in nature. We also showed that Morris water maze data have a compositional structure \cite{tml2025}. Extensions of this method can be developed for ordinal group membership, such as an extension of the Jonckheere-Terpstra test for compositional data. One could conceive longitudinal models for compositional data, where each time point is represented by a composition rather than a numeric value. The baseball data is one such example where it would be of interest to track the changes in branch proportions through time and develop visualizations to show the changes in proportions through the seasons. Furthermore, we can extend our model by adding covariates, as in the case of Dirichlet regression. 

Studies of Type I error and power for the LRT developed here show that at least 100 observations are necessary to have decent power to detect a difference \cite{tml2025}. This is due to the asymptotic nature of the LRT. There might be other testing frameworks than the LRT which would yield more powerful tests. However, the need for relatively large samples is somewhat moot in this age of big data, particularly considering the amount of data that is gathered for sports and other disciplines on a daily basis. Our results demonstrate LDM is valuable methodology for determining differences among separate populations where the observations are compositional in nature.

\section*{Acknowledgements}
The authors thank Brad Null for providing his data on at-bat outcomes for all MLB players from 2000-2010. We also thank SMU for providing a University PhD Fellowship to Bianca Luedeker.

\bibliographystyle{plain}
\bibliography{SDSSbib}

\end{document}